\documentclass[conference]{IEEEtran}
\IEEEoverridecommandlockouts
\usepackage{subcaption}
\usepackage{cite}
\usepackage{amsmath,amssymb,amsfonts,amsthm}
\usepackage{algorithm}
\usepackage[noend]{algorithmic}
\usepackage{multicol}
\usepackage{graphicx}
\usepackage{textcomp}
\usepackage{xcolor}
\usepackage{booktabs}
\usepackage{mathrsfs}

\usepackage{caption}
\usepackage{subcaption}

\def\BibTeX{{\rm B\kern-.05em{\sc i\kern-.025em b}\kern-.08em
		T\kern-.1667em\lower.7ex\hbox{E}\kern-.125emX}}
\begin{document}

	\title{Secure Decentralized Learning with Blockchain
	}
	
	\author{\IEEEauthorblockN{Xiaoxue Zhang*, Yifan Hua*, and Chen Qian}
		\IEEEauthorblockA{
			\textit{University of California Santa Cruz}\\
			xzhan330@ucsc.edu, yhua294@ucsc.edu and qian@ucsc.edu}
            \thanks{*Zhang and Hua made equal contributions to this work.}
	}
	\maketitle
	\pagestyle{plain}

	\begin{abstract}
		Federated Learning (FL) is a well-known paradigm of distributed machine learning on mobile and IoT devices, which preserves data privacy and optimizes communication efficiency. To avoid the single point of failure problem in FL, decentralized federated learning (DFL) has been proposed to use peer-to-peer communication for model aggregation, which has been considered an attractive solution for machine learning tasks on distributed personal devices. However, this process is vulnerable to attackers who share false models and data. If there exists a group of malicious clients, they might harm the performance of the model by carrying out a poisoning attack. In addition, in DFL, clients often lack the incentives to contribute their computing powers to do model training. In this paper, we proposed Blockchain-based Decentralized Federated Learning (BDFL), which leverages a blockchain for decentralized model verification and auditing. BDFL includes an auditor committee for model verification, an incentive mechanism to encourage the participation of clients, a reputation model to evaluate the trustworthiness of clients, and a protocol suite for dynamic network updates. Evaluation results show that, with the reputation mechanism, BDFL achieves fast model convergence and high accuracy on real datasets even if there exist 30\% malicious clients in the system.
	\end{abstract}

\section{Introduction}

Federated Learning (FL)~\cite{mcmahan2017communication} is a distributed machine learning (ML) paradigm that allows training ML models across numerous distributed devices, such as mobile and IoT devices. Those edge devices hold their data locally and collaboratively perform training tasks without directly sharing training data among them to ensure privacy. The trained ML models are then aggregated on a central server, called the aggregator.
The aggregator first distributes a global model to clients. 
Each client trains the model locally using its own data and generates a model update, which is then sent back to the aggregator. The aggregator aggregates these updates to update the global model, and distribute it to clients for further training. FL preserves data privacy by enabling decentralized model training~\cite{ma2020safeguarding, wei2020federated}, saving communication costs by avoiding moving raw data, and reducing computational costs by leveraging the computing resource of each device.
However, the existence of the centralized aggregator makes  FL vulnerable to a single point of failure~\cite{li2020federated}. Once the centralized aggregator is compromised, the whole FL system will fail. Also, the aggregator that frequently exchanges models with clients can become the bottleneck of the system. 

The recently proposed concept of Decentralized federated learning (DFL)~\cite{he2018cola, sun2022decentralized,Liu2023AAFL,liao2023ADFL} provides a solution for aforementioned problems by removing the involvement of the central server. In a DFL system, instead of communicating with a central aggregator, clients directly exchange model updates with a subset of other clients, also known as their ``neighbors", using P2P communication. Clients keep exchanging model updates until their local models converge to a model that reflects the features of data from all clients. Thus, DFL improves the limitations of having a single point of failure, trust dependencies, and bottlenecks on the server side in the traditional FL.
However, DFL still has some challenges such as malicious clients, low-quality models, and the lack of incentives, which undermines the reliability of the whole system. Given the large number of participants in the DFL system, it is unrealistic to simply assume all the clients are honest and follow the protocols to do the training correctly. Therefore, there may exist malicious clients sharing false model updates about their local training results. Also, some clients with low-quality models might also affect the performance of their neighbors with high-quality models, and these errors may be further propagated in the whole network. Besides, how to motivate data owners to participate in the system and continuously contribute their data to the FL model remains a challenge.

\begin{figure}[t]
	\centering
	\includegraphics[width=0.4\textwidth]{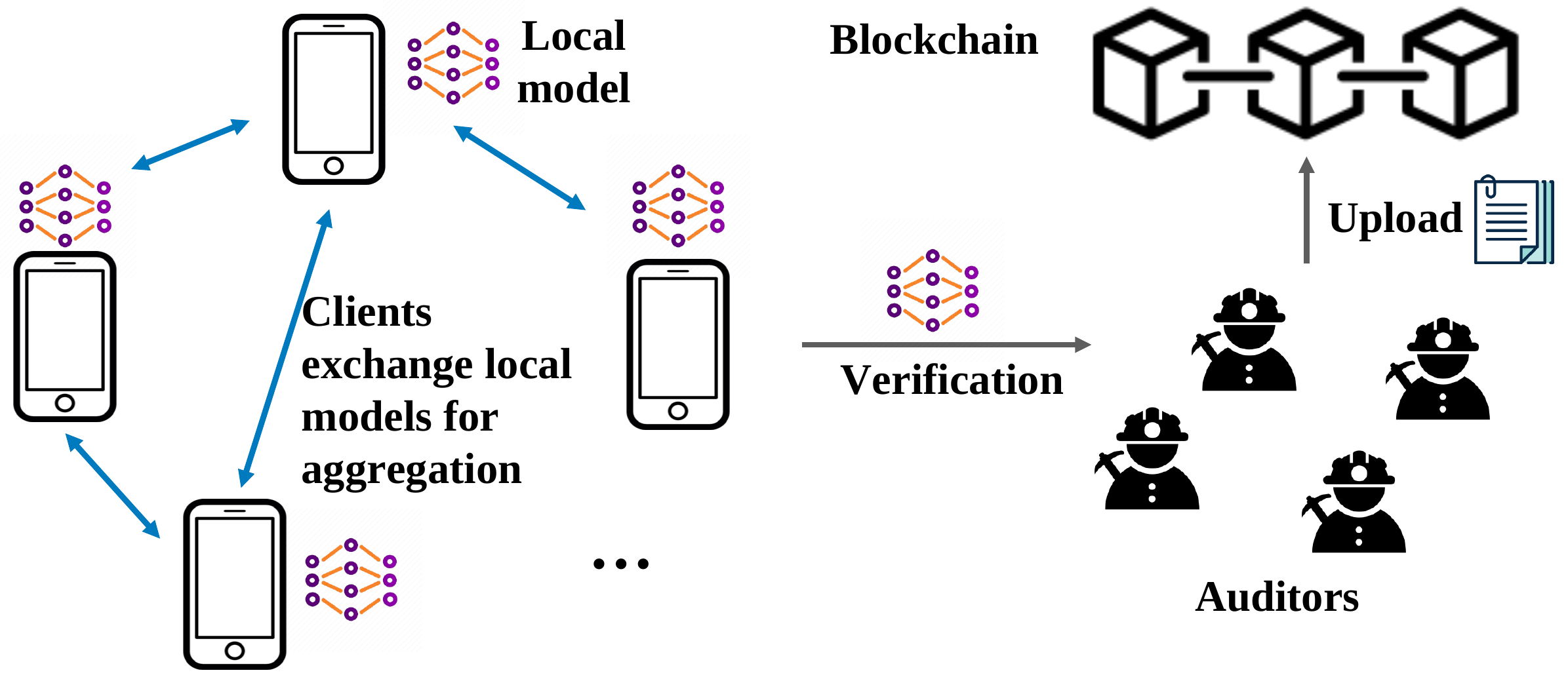}
	\caption{Blockchain-based Decentralized Federated Learning: clients form a peer-to-peer overlay network to exchange models, and auditors are responsible for model verification.}
 \vspace{-3ex}
	\label{fig:BDFL}\vspace{-1ex}
\end{figure}

Blockchain, as a distributed ledger technology built on a peer-to-peer network, provides a possible solution for the security and incentive issue in the DFL system~\cite{shayan2020biscotti}, as shown in Fig.~\ref{fig:BDFL}. It provides trust by allowing all participants to verify transactions submitted to the blockchain with its underlying provable cryptography and consensus protocol. Every participant could verify each model update before it can be aggregated and confirmed in the blockchain. However, directly storing all model updates on the blockchain is not feasible due to the significant costs incurred by data storage and computation. And pushing large model data to the blockchain has the problem of heavy latency, limited block size, and transaction size. Thus, we introduce an auditor committee and the reputation mechanism for model verification. Auditors are responsible for managing the clients' reputations according to the quality of their model updates. Instead of models or gradients, auditors will only put the clients' reputations on the blockchain and update periodically.

In this paper, we introduce BDFL, a Blockchain-based Decentralized Federated Learning system with an incentive mechanism and reputation model. We introduce a set of auditors for model verification, and honest clients can receive incentives while clients providing malicious models will be punished. In our reputation model, clients' reputation scores are assigned by auditors according to their model verification results, which will be updated periodically. Clients with higher reputations will have a higher probability for their model updates to be accepted by neighbors, and thus gain more profits from the system.



In summary, this paper makes the following contributions:
\begin{itemize}
	\item We design BDFL, the first blockchain-based fully decentralized federated learning system for model verification with high learning accuracy and system robustness.
        \item We design and implement the BDFL protocol suite. We introduce an incentive mechanism to encourage clients to participate in the model exchange, and a reputation model to evaluate the trustworthiness of each client to avoid malicious model updates from attackers.
        \item We evaluate BDFL using experiments on real ML datasets. We find that BDFL achieves a high model accuracy and fast convergence. It also has strong resilience to client dynamic and malicious model updates.
\end{itemize}

The rest of this paper is organized as follows. 
The system overview and model are presented in Section~\ref{sec:overview}.
We describe a protocol overview in Section \ref{sec:design} and the detailed design of the BDFL in Section \ref{sec:protocol}. Section~\ref{sec:evaluation} presents the evaluation results of our protocol. Section~\ref{sec:related} describes the related work. Section~\ref{sec:conclusion} concludes this work.

\section{Overview}
\label{sec:overview}

\addtolength{\textheight}{0.01in}

\subsection{Network Model}
We consider a decentralized learning system with a large number of clients, in which clients can join or leave anytime. Those clients, such as edge devices, are willing to train models using their own data locally, and exchange model updates with their neighbors to get some profits. 
There are also some auditors in the system who are responsible for model verification. They are a group of nodes that have read and write access to the blockchain, similar to the miners in common blockchain systems. They work as a committee to verify the model updates using public validation data in the system, and generate new blocks of reputation information. Honest auditors will receive rewards for correct verification, and dishonest ones can get punishment if detected.

Our network model is divided into two parts, the decentralized federated learning network, and the blockchain network. We model the BDFL network as an undirected graph $G = (V, E)$, where $V$ is the set of clients, and each link $e=(u, v) \in E$ represents that client $u$ and $v$ are neighbors, and can directly exchange local ML models. We assume clients have equal roles in the BDFL network and similar numbers of neighbors. Clients have read access to the blockchain, and can communicate with auditors for model verification. 
In the blockchain network, the clients' and auditors' identities and model verification information are recorded in the blockchain in the form of transactions. To encourage the participation of more users, auditors successfully performing verification and clients honestly providing model updates will receive incentives, which are guaranteed by smart contracts in the blockchain. Malicious participants can also be identified by the blockchain to protect the quality of the overall model.

\subsection{Blockchain Model and Assumptions}
The blockchain in the BDFL system should support smart contracts, which are responsible for managing client reputation and ensuring auditors behave appropriately. Both clients and auditors need to register on the blockchain first to join the BDFL system. 
Since we use the blockchain as the underlying root of trust, if it is compromised by an adversary, the correct functionality of the BDFL system cannot be guaranteed. Therefore, we assume that the proportion of consensus participants corrupted by an adversary for the blockchain is bounded by a threshold to ensure safety and liveness for the underlying blockchains. Following Byzantine fault-tolerant settings, we assume the proportion of adversaries is less than 33\% of the total number of consensus participants~\cite{gervais2016security}. 

\subsection{Attacker Model}
\vspace{-1ex}
We assume the attackers can potentially gain physical access to some clients in which the data and model are stored, and complete control of their network connections. They may want to destroy the global model by performing poisoning attacks. They will train models using false data and exchanging adversarial updates with neighbors~\cite{biggio2012poisoning, fung2020limitations}. They may also delay or prevent the client they control from accessing the blockchain for an unbounded amount of time. 
They are curious about clients’ private information, and can perform information leakage attacks by observing the model updates, and then recover details about clients' training data~\cite{hitaj2017deep, salem2018ml}. To prevent such attacks during model update transmission, clients can send differentially-private updates to mask their gradient~\cite{abadi2016deep, dwork2014algorithmic}.
We assume most of the auditors will correctly follow the protocol, 
and part of them may be corrupted in a Sybil attack. But attackers cannot control more than 1/3 of total auditors according to Byzantine fault tolerance.
We assume that all clients in the BDFL system could securely conduct initialization, in which they can correctly obtain the first version of the global model.
Auditors have relatively equivalent computation resource~\cite{zamani2018rapidchain}, and rational public validation data to perform model update verification.

\subsection{Requirements}

\textbf{Security:}
The main security requirement of the BDFL system is that it should enable model update exchanges between clients safely and correctly. We consider the security of both clients and auditors. Honest clients who provide the correct model updates will be acknowledged with profits and gain a better reputation, while clients with malicious model updates will ultimately be detected, and has a dramatic drop in their reputation. Whenever a client receives a model update from their neighbor, they can verify the correctness of the update with the help of auditors and the blockchain. If the model update successfully passes the verification, the client will accept it to further aggregate with their own model locally. If the model is considered to be malicious, the client will reject this update. As for the auditors, if an auditor claims an incorrect model update could pass verification, this behavior will certainly be detected within the blockchain, as it requires the approval of more than 2/3 of the auditors for the verification result to be confirmed in the blockchain. The malicious auditor will eventually lose all their collateral as ensured by the smart contract.

\textbf{Auditability:}
Any clients with read access to blockchain are able to get the latest reputation of all clients. Clients can also audit the model updates from their neighbors with the help of auditors and the blockchain.

\textbf{Privacy:}
The BDFL system should be able to keep client training data private by preventing information leakage attacks. The auditors received masked model updates from clients can successfully verify the correctness of the model, but cannot learn any information on the clients' training data.

\textbf{Robustness:}
In the BDFL system, the local models on the honest clients should eventually converge to a model that reflects the features of data from all clients with high accuracy.
The system should keep robust under attacks, which means, even if there exist attackers, 
the system should still achieve equivalent model accuracy. Moreover, the DFL network should be resilient to client dynamics such as client joins, leaves, and failures.

	\section{Design Overview}
\label{sec:design}

\addtolength{\textheight}{0.01in}

\begin{table*}[t]
	\centering
	\footnotesize
	
	{\caption{BDFL API}
		\label{table:API}
		\vspace{-1ex}
	}
	{
		\begin{tabular}{l l l l}
			\hline
			\textbf{BDFL APIs} & \textbf{Inputs} & \textbf{Outputs} & \textbf{API Description} \\
			\hline
			\textit{Topology Maintenance:} \\
			$\mathtt{join\_network}$ & $u, v$ & $u_{id}, rep_u$ & Join the DFL network to find its correct neighbors and register in the BDFL system. \\
			$\mathtt{leave\_network}$ & $u, u_{id}$ & - & Terminate model exchange and leave the network. \\
                $\mathtt{maintenance}$ & $u$ & - & Maintain the correct DFL network topology by checking the liveness of all $u$'s neighbors. \\
			\hline
			\textit{Model Exchange:} \\
                $\mathtt{local\_verify}$ & $u, v, \omega_v$ & Boolen & $u$ locally pre-evaluate the accuracy of model update $\omega_v$ from their neighbor $v$. \\
			$\mathtt{request\_verify}$ & $u, v, \omega_v$ & $\sigma_{\omega_v}$ & $u$ request model verification on the model update $\omega_v$ from their neighbor $v$. \\
                $\mathtt{aggregate\_model}$ & $\omega_u, \omega_v$ & $\omega'$ & $u$ locally aggregates the models from their neighbors. \\
			\hline
		\end{tabular}
	}
	\vspace{-2ex}
\end{table*}

In a fully decentralized overlay network for DFL, the BDFL protocol suite provides two sets of protocols for clients: 1) a DFL network \textit{Topology Maintenance Protocol} to build the overlay network and recover it from churn; 2) a \textit{Model Exchange Protocol} which includes model verification to achieve fast model convergence for heterogeneous clients and asynchronous communication. Table~\ref{table:API} shows the API that BDFL provides to clients. 
BDFL generates unique identifiers and an initialized reputation for each client, e.g., when a client $u$ joins the DFL network and registers in the BDFL system for model exchange, a unique identifier $u_{id}$ is returned as a handle to be used in reputation management and subsequent API calls.

The BDFL protocols work with any overlay topology and we apply a recently proposed overlay topology as a case to study BDFL~\cite{hua2022efficient}, which is based on near-random regular graphs (RRGs)~\cite{yu2014space}.


In P2P model exchanges, a client with low-quality local models might pollute its neighbors with high-quality models. This could lead to further propagation of these errors throughout the overlay network. Thus, every time when a client receives a model update from its neighbor, the client will first self-evaluate the confidence of this model. If the reputation of this neighbor is too low, the client can directly reject this model update. If the client feels the model is unreliable, the Model Exchange Protocol allows them to request model verification from auditors. Auditors then use an anonymized public validation dataset to do the model verification~\cite{sheller2020federated}. If the computed accuracy by the auditors drops a lot compared to its previous model accuracy, this model update is considered to fail the verification. Auditors will announce the verification result to the corresponding client, and reduce the reputation of the client who provided this model update. Otherwise, the client performs the model aggregation locally after receiving the correct verification result from auditors.

\textbf{DFL Topology.}
In BDFL, each client is identified by a set of \textit{virtual} coordinates $C$, which is an $L$-dimensional vector $<x_1,x_2,...,x_L>$. Each element $x_l$ is a random real number computed as $H(IP_x|i)$ where $H$ is a publicly known hash function and  $IP_x$ is $x$'s IP address. 
We create $L$ virtual ring spaces~\cite{yu2014space} such that each client in the $i$-th ring space is virtually positioned based on its coordinate $x_i$. In each virtual ring space, every client has two adjacent clients based on their coordinates, forming overlay neighbors for model exchanges. 
Each client can have a maximum of $2L$ neighbors, with $L$ serving as a trade-off parameter between communication and convergence. A larger $L$ leads to more model exchanges but also increases the communication cost.

\textbf{Auditors.} 
Auditors are groups of nodes that have read and write access to the blockchain. They work jointly with the blockchain for client registration, model verification, and reputation management. 
The system leverages a public smart contract (\textsf{aSC}) to maintain an auditor list and ensure the correct behavior of the auditors. They are required to lock some collateral to be registered with this smart contract, i.e., \textsf{aSC} can verify the auditor’s digital signature and knows the auditor’s public key.
We assume the majority of the auditors are reliable. They are willing to follow the protocol to get profits, and punish malicious auditors for misbehavior. For each client's model verification request and reputation update, a minimum number of auditors are required to sign the result before packing the update into the blockchains, thus tolerating a fixed percentage of auditors' failures to some degree. 
A common way to address such concerns is to use Byzantine-fault tolerant protocols~\cite{mavroudis2020snappy}. For example, the auditors could use a BFT consensus such as~\cite{castro2002practical} to stay up to date with all the coming requests from users. Such a solution can tolerate up to 1/3 faulty auditors.
Thus, in BDFL, the smart contract defines that at least 2/3 of the auditors are required to sign each verification or update request before sending them to the blockchain, and only the auditors who correctly sign the request can get rewards.


\textbf{Model Update.}
Different from FL, BDFL does not require a central server for model aggregation. Instead, every client can run the model aggregation locally using the model updates gathered from its neighbors. Once clients successfully prepare models locally, they can collect model updates from their neighbors for further aggregation. Clients always reject model updates from neighbors with low reputations. Clients then query auditors for model verification on the rest of the model updates. After verification by the auditors, clients run the model aggregation on all the correct model updates.

\textbf{Reputation.}
In BDFL, each client is assigned a reputation value by auditors which reflects its trustworthiness. Clients should have a higher possibility to accept model updates from honest clients, and reject those from malicious ones. To prevent poisoning attacks from malicious clients, every time when auditors detect a model update of low accuracy, the auditors will decrease the reputations of the corresponding misbehaving client. On the other hand, honest clients will gain a reputation increase by providing good model updates.

	\section{Protocol Design}
\label{sec:protocol}

This section describes the design of BDFL protocols.

\vspace{-1ex}
\subsection{Topology Maintenance}
The Topology Maintenance Protocol in BDFL system includes $\mathtt{join\_network}, \mathtt{leave\_network}$ and $\mathtt{maintenance}$ as shown in Table~\ref{table:API}. 

\textbf{Join.}
Assume we have a correct DFL network topology with $n$ clients currently. A new client $u$ now boots up and wants to join the BDFL system for future model exchange. Before joining the DFL network, $u$ has to know one existing client $v$ in the overlay. $u$ assigns itself a random coordinate in the virtual ring spaces as its position. Then it sends join requests to its neighbor $v$, and tries to find all its neighbors in the network. To achieve this, $u$ lets $v$ send a $Neighbor\_discovery$ message which includes $u$'s IP address to the current DFL network using greedy routing to $u$'s location in each ring space respectively. $Neighbor\_discovery$ stops at the client $w$ who is closest to $u$. In each virtual ring space, $w$ finds the adjacent node $p$ from its two adjacent nodes to insert $v$ in between according to $u$'s coordinate.

\textbf{Leave.}
When a client wants to leave the system, $\mathtt{leave\_network}$ should guarantee that the BDFL system can still maintain a correct DFL network topology.
Assume client $u$ wants to leave by running $\mathtt{leave\_network}$. $u$ sends messages to its two adjacent clients in each virtual ring space, and tells them to add each other to their neighbor sets. 

\subsection{System Maintenance}
Auditors are responsible to help maintain the BDFL system. They will track client information such as client identity and reputation. They will update this information to the blockchain periodically, and all the clients can easily check this information by reading the blockchain. To do this, each auditor maintains a local table to record client information. It includes three parts: the reputations of all the valid clients, which are updated according to the quality of models provided by clients; a joining client set, which is the clients who join the system after the last update; and a leaving client set. This leaving client set includes two kinds of clients. One is the clients who want to stop exchanging models with others and leave the system. The other is the clients who have been inactive for a long period, or with a very low reputation. The system will kick them out by adding them to the leaving client set.

The BDFL system should also be able to maintain a correct DFL network topology experiencing client failures. The $\mathtt{maintenance}$ protocol requires every client to send neighbors a heartbeat message periodically, to filter out inactive clients. 

Clients join BDFL to collaboratively train ML models. A new client joins the BDFL system by calling the $\mathtt{join\_network}$ function. In this function, in addition to the \textbf{Join} process to the network topology, as previously described, the client also needs to register on the BDFL blockchain to participate in future model exchanges. 
To do this, the new client, denoted as $u$, sends a join query to auditors. Auditors will record the client's information $u_{id}$ and assign a default reputation $rep_u$ to it. The auditor committee will pack the client information update to the blockchain periodically. Once the update message which includes the new client $u$ is confirmed on the blockchain, $u$ is then considered to have successfully registered within the BDFL system, and can start model training. 
Currently, $u$ has no model, and has to initialize the training process by first gathering models from its neighbors. $u$ has to verify the correctness of the model with the help of auditors before aggregating them. After getting verification results from auditors, $u$ only chooses to use the correct models from the neighboring clients, and discard the others.
$u$ then locally generates an aggregated model as its own model. And later, it will keep gathering model updates from its neighbors and continue updating its local model to improve accuracy in the future. It will also exchange its local training model with its neighbors to contribute to the whole system and get profits.

Clients exchange local models with neighbors periodically, and the models will be evaluated by auditors which will affect their reputations.
Thus, client reputations are updated by auditors during model verification. If a malicious model update is detected and verified by auditors, the corresponding client who provides this model will be punished with a low reputation. Clients with very low reputations will be removed from the BDFL system forever. To achieve this, auditors check their local client table periodically to filter out the clients with low reputations, and add them to the leaving client set. After this client table is confirmed in the blockchain, those clients are removed from the system successfully.

\begin{figure}[t]
	\centering
	\includegraphics[width=0.45\textwidth]{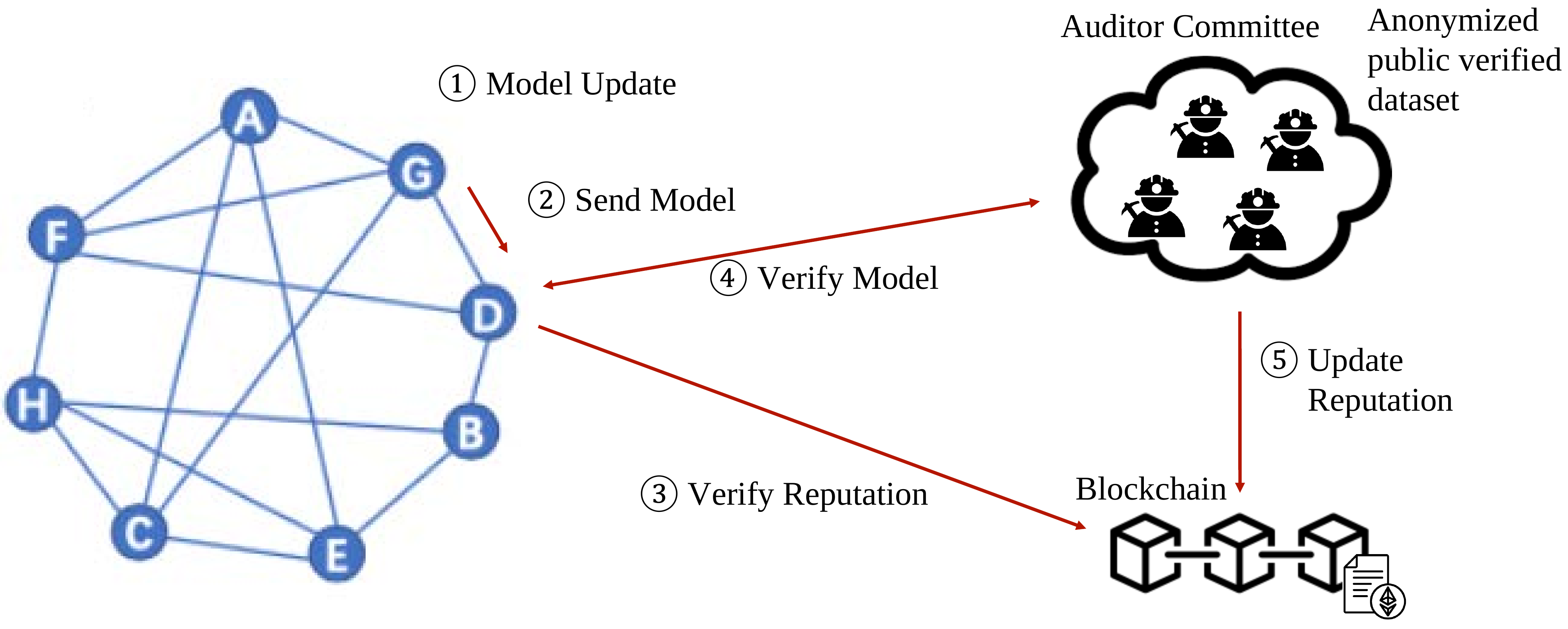}
	\caption{Protocol overview of BDFL.}
 \vspace{-2ex}
	\label{fig:overview} \vspace{-2ex}
\end{figure}

\subsection{Model Exchange}
In the model exchange, each client exchanges its local update with neighbors to do aggregation. However, clients might be not willing to send their models directly in concern that others might learn about their dataset information. It has long been established that gradients often leak sensitive information about clients’ local datasets~\cite{bell2020secure, bonawitz2017practical}, and therefore, it is necessary for clients to hide their model updates to keep privacy. 
To prevent information leakage attacks, clients use differential privacy (DP) to hide their updates during model exchange and verification by adding noise sampled from a normal distribution~\cite{dwork2014algorithmic}. We follow the concept of $(\epsilon,\delta)$ differential privacy as being applied in many previous works~\cite{shayan2020biscotti, wei2020federated}. $(\epsilon,\delta)$-DP provides a strong criterion for privacy-preserving of distributed data processing systems. Thus, each client constructs a noisy model update by adding some noise to the gradient such that $\omega^t = \zeta^t + \omega_{original}^t$.

Another challenge of performing decentralized model exchanges is that there is no central server to evaluate the quality of models from different clients. Thus, our Model Exchange Protocol is designed to validate the model update to mitigate the impact of malicious models.
Fig.~\ref{fig:overview} illustrates the overview of the Model Exchange Protocol. It runs on both the clients and the auditors. Every client prepares its model locally, and exchanges the model update with neighbors once they finish training. BDFL uses asynchronous communication and allows each client to use a different communication and training time period in a round. 
Assume an honest client $u$ has a local model $\omega_u^t$ in its current round $t$. And it has three neighbors $v, w$ and $x$ to exchange models with. 
In the neighbor set of $u$, in addition to the IP addresses and coordinates, it also stores the fingerprint $f$ of the most recent model updates received from each neighbor, computed by hashing the model updates by a public hash function. 
When a client successfully prepares a model update to send, its neighbors would check the fingerprint $f$ to avoid repetitive updates. This approach effectively mitigates unnecessary traffic, thereby reducing the frequency of exchanging duplicate models.

At client $u$, after gathering model exchanges $\omega_v^t$, $\omega_w^t$, and $\omega_x^t$ from all its neighbors, $u$ first checks their reputations by reading the blockchain, and rejects the model updates from neighbors with low reputation. Let's say client $x$ has a low reputation. $u$ will directly reject the model update $\omega_x^t$, and only continue to verify the correctness of the remaining model updates for aggregation. Now $u$ needs to verify model updates $\omega_v^t$ and $\omega_w^t$ by sending verification requests $request\_verify(u, v, \omega_v^t)$ and $request\_verify(u, w, \omega_w^t)$ to auditors. Auditors use the public validation data to verify the model updates. The smart contract maintains a history of model verification results, which include an average model accuracy $\mu_{t-1}$ and standard deviation $\sigma_{t-1}$ of the latest 20 epochs. If the computed accuracy of the received model update, e.g. $\omega_w^t$, is less than $\mu_{t-1} - 2\sigma_{t-1}$, this model update is considered to be malicious. Auditors will reply with the verification results to $u$, and punish $w$ of the malicious update by decreasing its reputation. 
For models that pass the verification, auditors will increase the reputations of the corresponding clients, and the smart contract needs to update the average model accuracy $\mu_{t}$ and standard deviation $\sigma_{t}$ as well. After receiving verification results from auditors, $u$ finds that $\omega_v^t$ is the only valid model update. $u$ will discard other model updates and later aggregate $\omega_v^t$ to its local model $\omega_u^t$ to train a new local model $\omega_u^{t+1}$.

In a more complex scenario, client $u$ discovers that it has received multiple valid model updates from various neighbors with different reputations after conducting model verification. $u$ always tends to believe clients with high reputations will provide model updates of high quality, while low-reputation clients might provide low-quality models. Even if they are all valid updates, $u$ still wants to limit the impact of low-quality models and amplify the impact of high-quality models in the aggregation. Thus, unlike the assumption that all model updates contribute equally to model aggregation, we let high-reputation clients have higher impacts. $u$ defines a set of confidence parameters $c$ for each model update, assigning higher $c$ values to clients with better reputations. The models from $u$’s neighbors are then aggregated as follows:
\begin{equation*}
    \omega_u^{t+1} = \frac{\sum_{j\in N \cup \{u\}} c^j\omega_j^t}{\sum_{j\in N \cup \{u\}} c^j}
\end{equation*}
, with $N$ being the neighbor set of client $u$. The model aggregation will be computed once every round and the models from each neighbor are always the most updated ones. In this way, clients with low reputations will have less impact on other clients.

\subsection{Model Verification}
In FL, every network provides a centralized validation dataset to validate the global model. The idea of using an anonymized public validation dataset for FL model validation is well adopted in the existing research~\cite{sheller2020federated,choudhury2020anonymizing}, and also used for detecting model poisoning attack~\cite{fang2020local}. We follow this assumption and adopt this idea for BDFL model update verification. 
We assume auditors have a public validation dataset $D$ contributed by their clients and a global model $\omega_0$. Before starting training, auditors collect training dataset samples from clients. We adopt $\epsilon$-differential privacy~\cite{dwork2006our} for protecting the data privacy of clients.
After gathering all those anonymous datasets, auditors pre-evaluate their quality by comparing the accuracy of $\omega_0$ on the public dataset $D$ and each fetch dataset $D_i$. If the accuracy of $D_i$ is much lower than the training result on $D$, auditors will reject this dataset $D_i$. Auditors then integrate the satisfied datasets into their local validation dataset to compose a new validation dataset $D$.
During the training process, clients continue collecting local data, and there might be new clients joining the system. Thus, auditors will gather clients' datasets periodically to update their validation datasets $D$.

When receiving a model verification request, auditors first check the fingerprint $f$. Then they can directly reply to the client with the previous verification result. Otherwise, auditors compute the accuracy by executing the model $\omega$ on the public validation dataset $D$. The smart contract \textsf{aSC} collects verification results from all auditors. It disregards those results that significantly deviate from the majority of the results, and considers auditors providing such results as malicious. \textsf{aSC} computes the average of the remaining accuracy results $A_t^\omega$, and compares with the previously stored average model accuracy $\mu_{t-1}$ and standard deviation $\sigma_{t-1}$ of the latest 20 epochs. If $A_t^\omega < \mu_{t-1} - 2\sigma_{t-1}$, $\omega$ will be judged as a malicious model update.

\subsection{Reputation Management}

Clients joining the system for the first time will be assigned an initial reputation $rep$. All the reputations are updated by auditors during each model verification. 
Let's say a client submits a model verification request on model update $\omega_u$ from its neighbor $u$. Auditors validate $\omega_u$ with public test data~\cite{peng2021vfchain} and get the corresponding accuracy values. \textsf{aSC} filters out those outlier results and uses the average of the remaining results as the accuracy value $A_t^{\omega_u}$ of model update $\omega_u$.
Note that \textsf{aSC} maintains an accuracy history of previous model updates, $\mu_{t-1}$ and $\sigma_{t-1}$, which are average model accuracy and standard deviation of the latest 20 epochs respectively. If $A_t^{\omega_u} < \mu_{t-1} - 2\sigma_{t-1}$, auditors will consider this client $u$ to be malicious, and punish $u$ with a decrease in its reputation from $rep_u$ to $\frac{rep_u}{2}$.
For honest clients, their model update accuracy might not always increase due to the imperfect of their dataset. The system tolerates this small accuracy decrease and will not lower their reputations as long as their model accuracy is larger than $\mu_{t-1} - 2\sigma_{t-1}$. Their reputations still keep the same in this round.
Other honest clients can gain a reputation increase $\Delta rep$ according to their model accuracy, which is computed as $\frac{A-\mu_{t-1}}{\sigma_{t-1}}\cdot 0.01$. For example, if the model accuracy of $\omega_u$ is $A_t^{\omega_u} = \mu_{t-1} + 2\sigma_{t-1}$, the reputation of the corresponding client $u$ will increase by 0.02.

If a malicious model update is detected and verified by auditors, the corresponding client will be punished with a decrease in its reputation by 50\%. However, a low reputation could result in its model update being declined by its neighbors in future model exchanges. Therefore, if such a client wants to continue participating in the model exchange and gain its reputation back, it must let auditors pre-verify its model updates prior to the model exchange period. If auditors validate this model update, the client will exchange its model along with the proof. Its neighbors can later confirm the correctness of this model update using the provided proof, thereby eliminating the need for another round of auditor verification.

Note that there is always a latency to post the new reputation (client table) to the blockchain. So in order to get the latest reputation information, clients can query auditors first. After some time, clients can verify the correctness of the query result by comparing it with the reputation confirmed in the blockchain. If clients detect that the reputation got from auditors is different from that in the blockchain, they can submit disputes to the blockchain for compensation, and the malicious auditors will get punishment and lose all their collateral.

\addtolength{\textheight}{0.03in}
\subsection{Incentives Mechanism}
The involvement of the blockchain always triggers some fees, such as running smart contracts and submitting transactions to the blockchains. Thus, auditors require some incentives in order to do model verification and reputation management. 
In BDFL, clients are the devices that want to train a better model using model updates from other clients in the system. If they directly aggregate models without verification, potential malicious models from attackers could lead to poor performance on their training models. Thus, we assume clients are willing to pay auditors for verification, and honest clients who provide them with a high-quality model update. Clients can also gain rewards by providing honest model updates to other clients.
Besides, there might be some services that want to use the final model directly without participating in the training process. Those services could request the model via the smart contract \textsf{aSC} by making a payment. \textsf{aSC} will then distribute the payment fee to auditors and clients according to their contribution.

An incentive mechanism is involved via the smart contract \textsf{aSC} to motivate participating clients and auditors to be honest and report the misbehavior.
Any device with read-and-write access to the blockchain is allowed to become an auditor by registering with the auditor smart contract (\textsf{aSC}) and providing collateral. They are required to verify model updates, and distribute incentives to honest clients while imposing punishment on misbehaving clients through the \textsf{aSC}. For each verification request, \textsf{aSC} gathers verification results from all auditors, and determines who provide correct verification results and record who always submit malicious results. If the number of incorrect verification results submitted by an auditor exceeds a predetermined threshold, this auditor will be judged as malicious and subject to penalties, such as losing all its collateral. It will then be removed from the auditor list in the smart contract. 
On the contrary, honest auditors who provide correct verification results will receive some rewards. 

With the verification result of a model update from a client, the smart contract \textsf{aSC} will determine if this client is honest and should get some rewards. For example, currently \textsf{aSC} has a verification result $A_t^{\omega_u}$ on the model update $\omega_u$ provided by the client $u$. \textsf{aSC} finds that it is a high-quality update with $A_t^{\omega_u}$ larger than the average model accuracy of the latest 20 epochs $\mu_{t-1}$, and $u$ should be rewarded with $c(A_t^{\omega_u}-\mu_{t-1})$.

In model verification, when receiving a model update with high accuracy, auditors could assume this model is good enough to be a global model that reflects the features of data from all clients correctly. And auditors can choose to maintain this model $\omega$ locally and send a digest $h(\omega)$ to \textsf{aSC}. \textsf{aSC} will record which auditors maintain a local copy of the global model, and the corresponding digest. This is to prevent auditors from manipulating a malicious global model.
If a service requests a model from \textsf{aSC} with a required fee, \textsf{aSC} will expose the auditor list who maintain the global model to the service. The service could randomly select an auditor to fetch the model and verify its correctness by querying \textsf{aSC}. \textsf{aSC} then distribute the fee collected from the service to both the auditor who provided the model and to all clients. The distribution to clients is based on their respective reputations, which reflect their individual contributions to the model.


	\section{Performance Evaluation}
\label{sec:evaluation}
\addtolength{\textheight}{0.03in}

\begin{figure*}[t]
    \centering
    \begin{subfigure}[b]{0.30\textwidth}
        \centering
        \includegraphics[width=\textwidth]{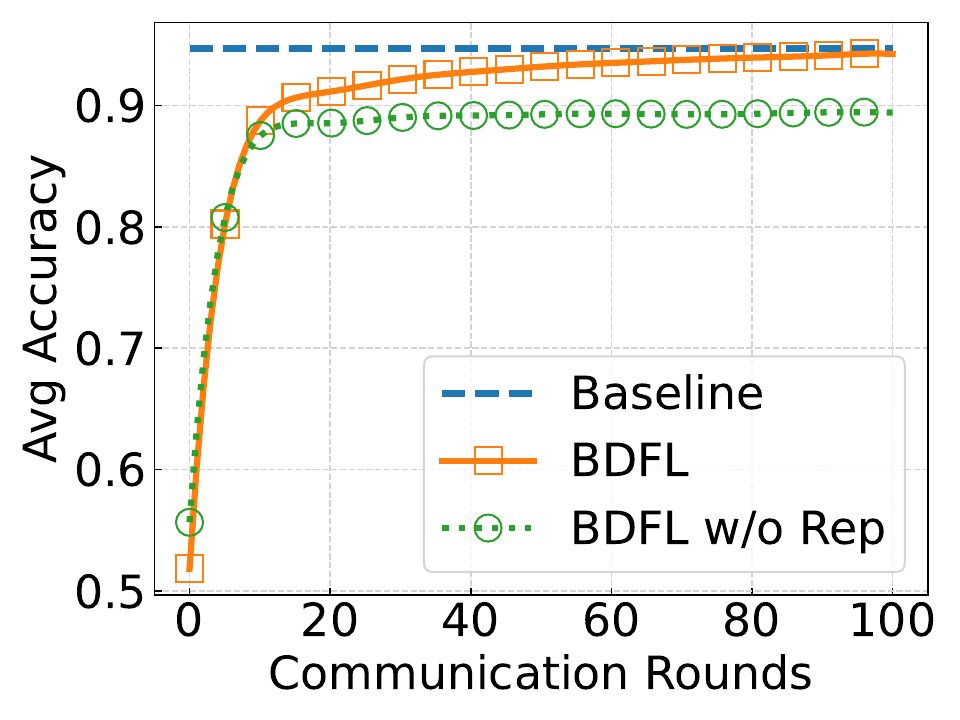}
        \vspace{-4ex}
        \caption{10\% Malicious Clients}
        \label{fig:acc-m-0}
    \end{subfigure}
    \begin{subfigure}[b]{0.30\textwidth}
        \centering
        \includegraphics[width=\textwidth]{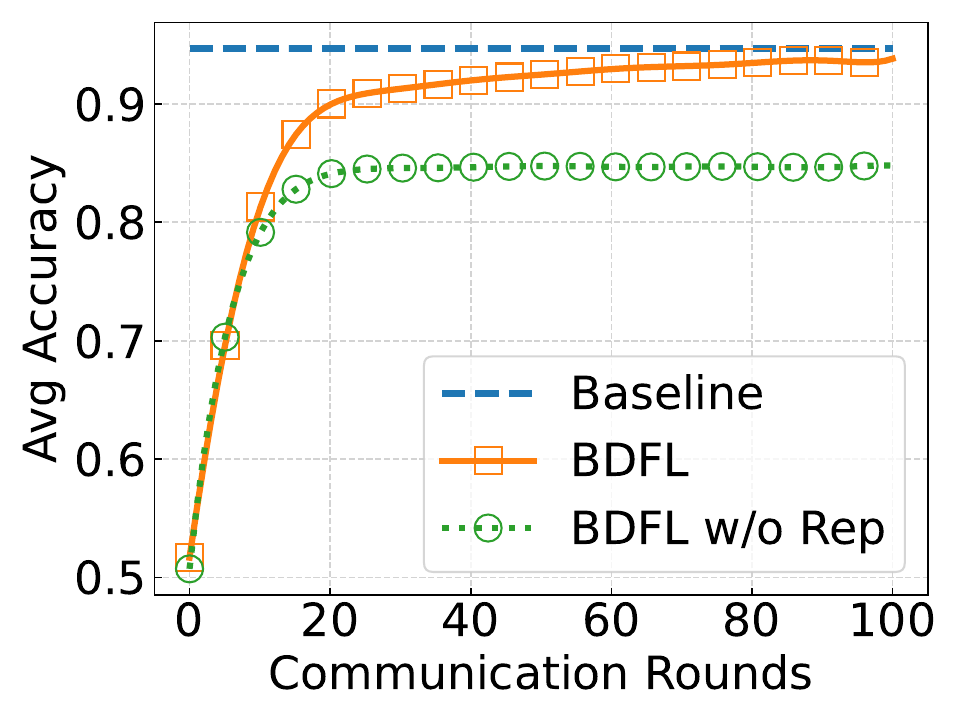}
        \vspace{-4ex}
        \caption{20\% Malicious Clients}
        \label{fig:acc-m-1}
    \end{subfigure}
    \begin{subfigure}[b]{0.30\textwidth}
        \centering
        \includegraphics[width=\textwidth]{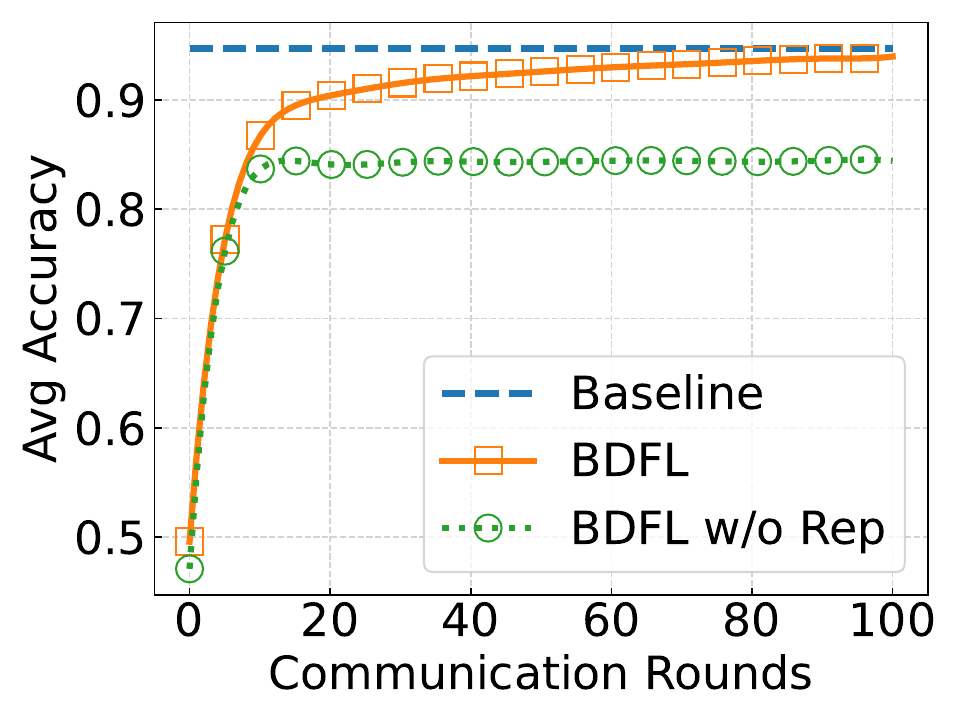}
        \vspace{-4ex}
        \caption{30\% Malicious Clients}
        \label{fig:acc-m-2}
    \end{subfigure}\vspace{-1ex}
   \caption{MNIST, Average Accuracy vs. Communication Rounds}\vspace{-2ex}
   \label{fig:acc-m}
\end{figure*}

\begin{figure*}
    \centering
    \begin{subfigure}[b]{0.30\textwidth}
        \centering
        \includegraphics[width=\textwidth]{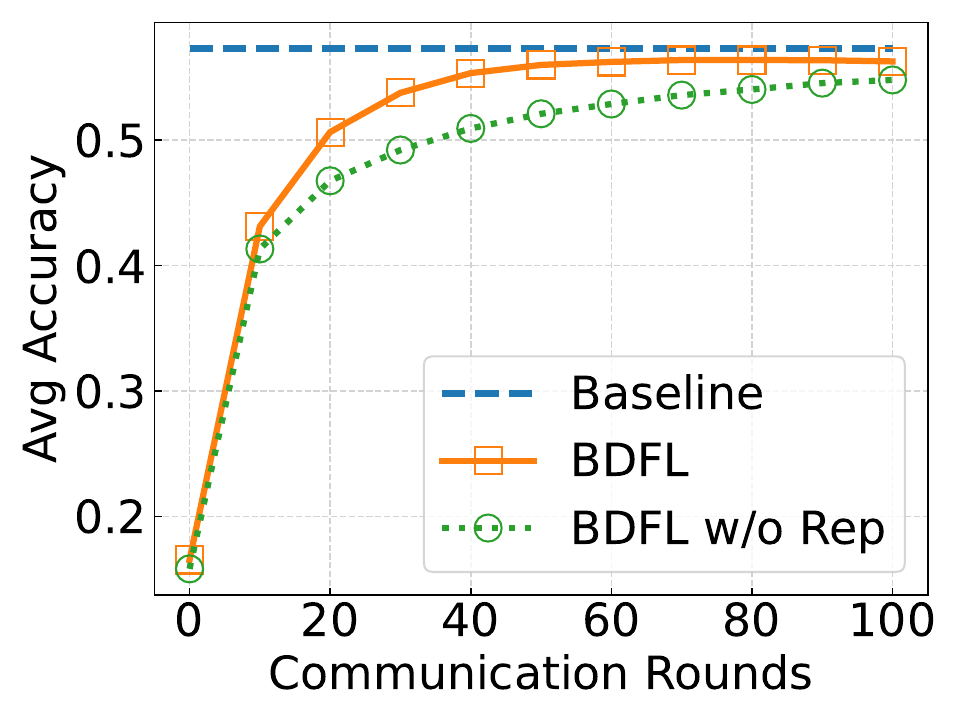}
         \vspace{-4ex}
        \caption{10\% Malicious Clients}
        \label{fig:acc-c-0}
    \end{subfigure}
    \begin{subfigure}[b]{0.30\textwidth}
        \centering
        \includegraphics[width=\textwidth]{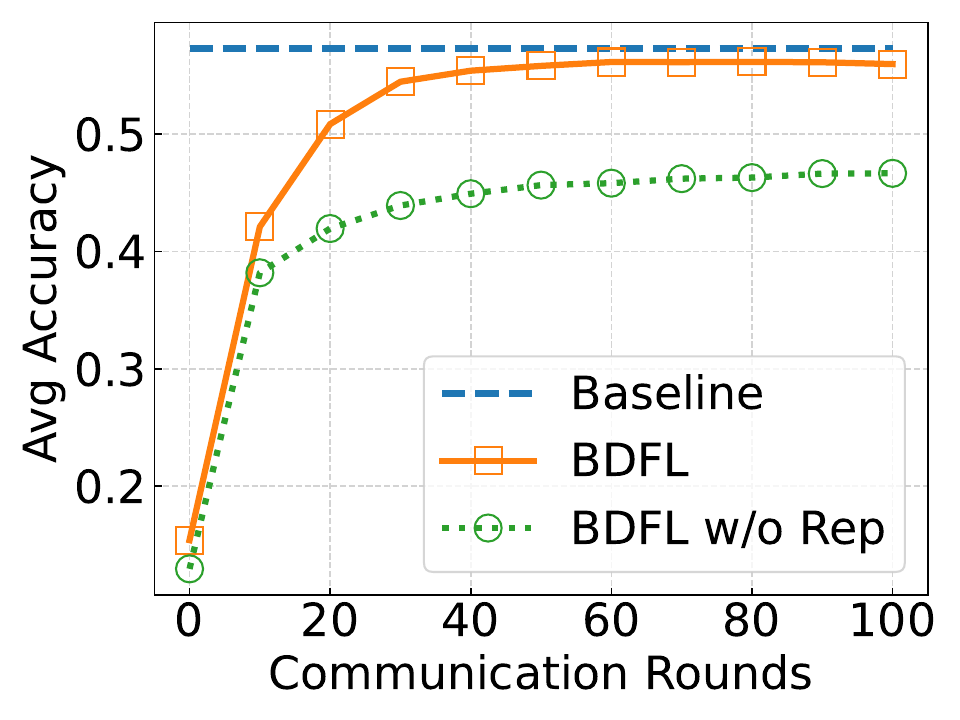}
         \vspace{-4ex}
        \caption{20\% Malicious Clients}
        \label{fig:acc-c-1}
    \end{subfigure}
    \begin{subfigure}[b]{0.30\textwidth}
        \centering
        \includegraphics[width=\textwidth]{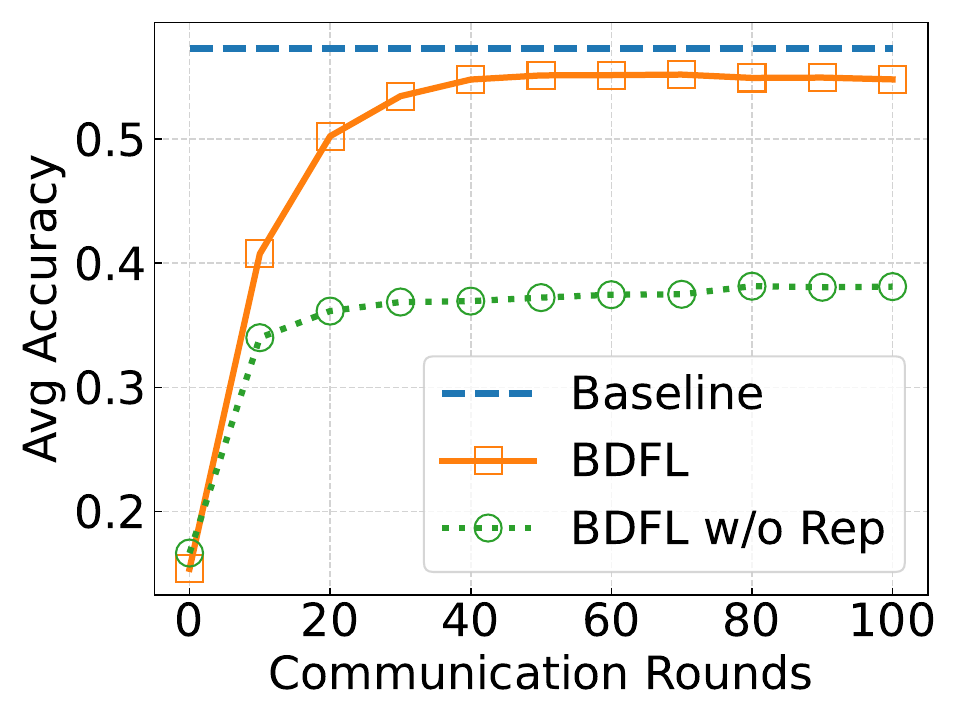}
         \vspace{-4ex}
        \caption{30\% Malicious Clients}
        \label{fig:acc-c-2}
    \end{subfigure}\vspace{-1ex}
   \caption{CIFAR-10, Average Accuracy vs. Communication Rounds}\vspace{-2ex}
   \label{fig:acc-c}
\end{figure*}

\begin{figure*}
    \centering
    \begin{subfigure}[b]{0.30\textwidth}
        \centering
        \includegraphics[width=\textwidth]{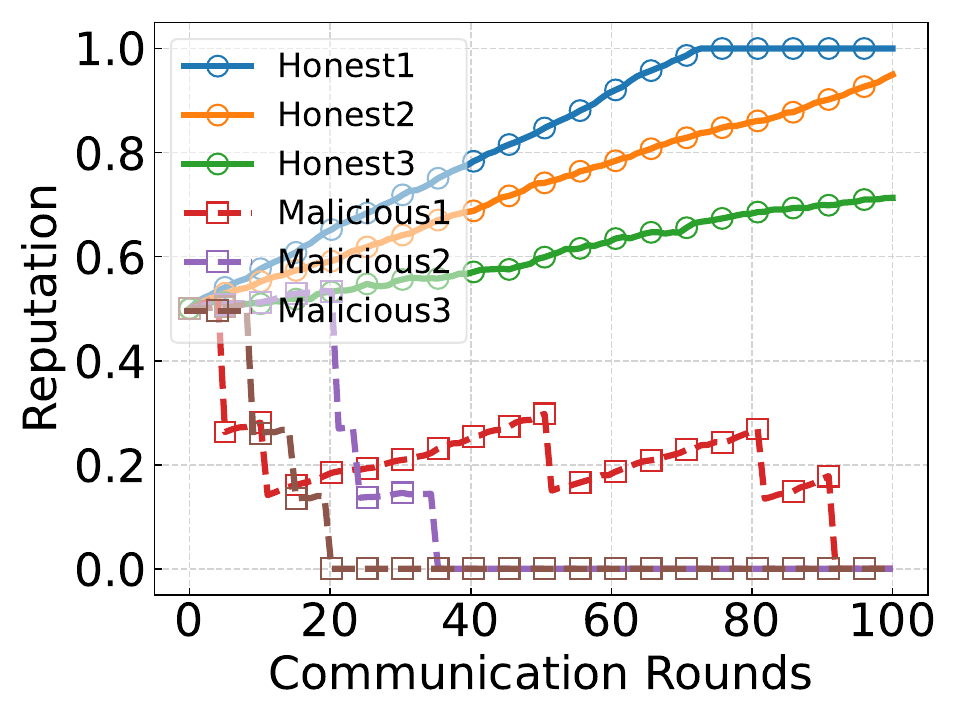}
        \vspace{-4ex}
        \caption{Reputation}
        \label{fig:re}
    \end{subfigure}
    \begin{subfigure}[b]{0.30\textwidth}
        \centering
        \includegraphics[width=\textwidth]{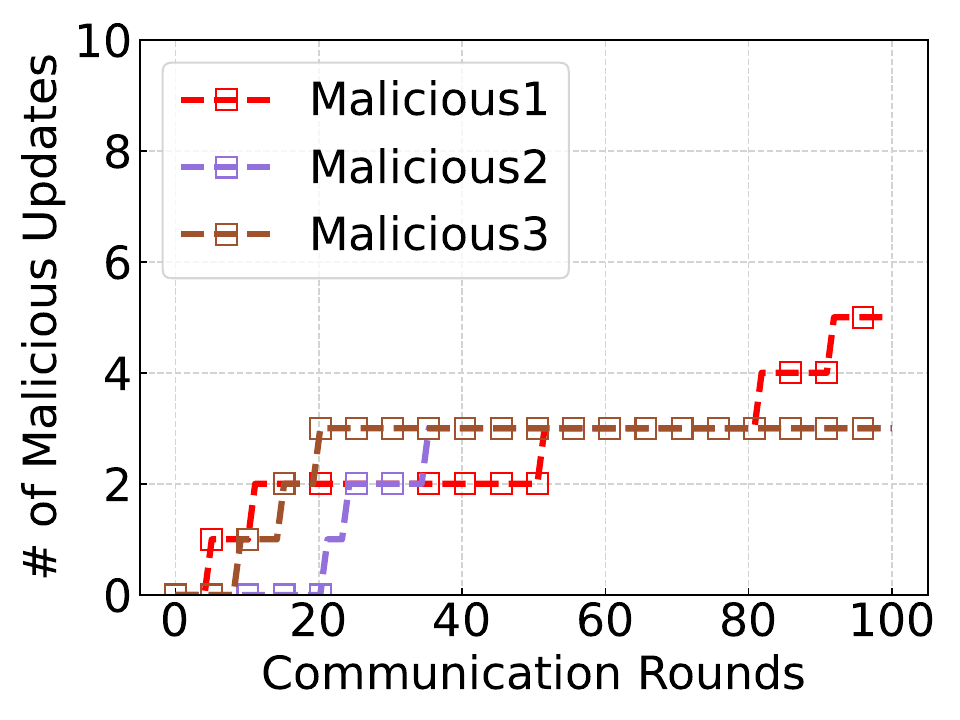}
        \vspace{-4ex}
        \caption{Malicious Updates}
        \label{fig:ma}
    \end{subfigure}
   \begin{subfigure}[b]{0.30\textwidth}
        \centering
        \includegraphics[width=\textwidth]{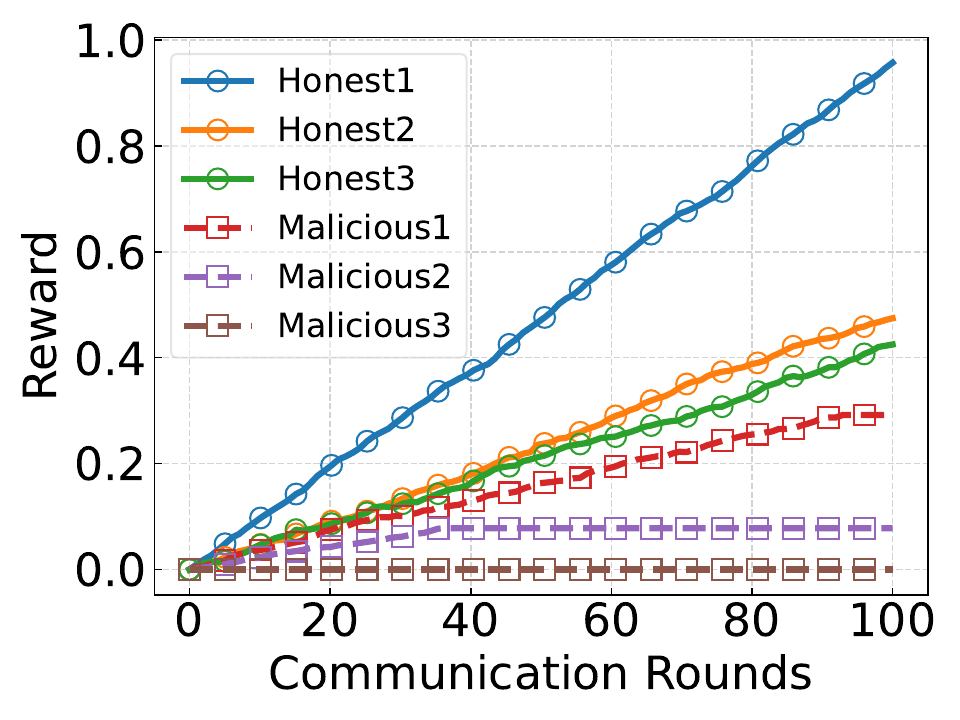}
        \vspace{-4ex}
        \caption{Relative Accumulative Reward}
        \label{fig:rew}
    \end{subfigure}\vspace{-1ex}
    \caption{MNIST, 20\% Malicious clients. We sample 3 honest clients and 3 malicious clients and plot their reputation, accumulative reward (relative to 'Honest1') and the number of successful malicious updates in the training period. Client 'Malicious1' performs attacks on rounds 4,10,50,80,90. Client 'Malicious2' performs attacks on rounds 8,14,19,34,52,54,71,77,81,90,95. Client 'Malicious3' performs attacks on rounds 20,23,34,58,61,83,95}\vspace{-2ex}
   \label{fig:det}
\end{figure*}


\subsection{Methodology}
We use a desktop machine with an NVIDIA GeForce RTX3060 Ti as the platform for our experiment. We build a 100-client network, use real data training, and simulate reputation management and incentives distribution. The purpose of this experiment is to test the performance of effectiveness of the reputation mechanism and robustness in a simulated environment. We choose MNIST~\cite{deng2012mnist} and CIFAR-10~\cite{krizhevsky2009learning} image classification as the task of BDFL. We use MultiLayer Perceptron networks (MLP) and Convolutional Neural Networks (CNN) respectively as the machine learning models.

\subsection{Evaluation Results}

\textbf{Robustness.}
We compare BDFL with the version without the reputation mechanism on two different datasets. The average accuracy of model updates verified by auditors is shown in Fig.~\ref{fig:acc-m} and Fig.~\ref{fig:acc-c} respectively. The Baseline shows the final model accuracy with no malicious clients in the system. In both two datasets, BDFL always demonstrates better accuracy under different proportions of malicious clients. The accuracy increase compared to the baseline without the reputation management mechanism is more obvious with a higher proportion of malicious clients.
Malicious clients can affect the average model accuracy by model poisoning attacks without verification. Honest clients will use those malicious models for aggregation directly, and thus harm the overall performance of the system. The more malicious clients in the system, the worse the average model accuracy. However, with the reputation management mechanism, malicious clients providing bad model updates can be detected by auditors and filtered out of the system. Malicious model updates will not be accepted by any client, and thus, even if there are 30\% malicious clients in the system, The average accuracy of BDFL with reputation mechanism only has a degradation of 0.73\% compared to baseline and BDFL without reputation mechanism has a degradation of 10.28\% for MNIST classification as shown in Fig.~\ref{fig:acc-m-2}. While in CIFAR-10 task, the average accuracy of BDFL with reputation mechanism has a degradation of 2.49\% and BDFL without reputation mechanism has a degradation of 19.20\% as shown in Fig.~\ref{fig:acc-c-2}.

\textbf{Reputation value evaluation.}
We monitor the reputation value changes for both honest and malicious clients in the MNIST dataset. The initial reputation of each client is set to be 0.5. The reputation value will be dynamically updated according to their behaviors. The percentage of malicious clients in this set of experiments is set to 20\%. Fig.~\ref{fig:re} shows the reputation changes for 3 randomly chosen honest clients and 3 randomly chosen malicious clients. We find that honest clients always gradually achieve a high reputation value, even if they might not always be able to provide model updates of high accuracy due to the imperfect of their datasets.
On the other hand, the reputation values of malicious clients will decrease to 0 in 100 rounds eventually, and they will be excluded from the system by auditors once extremely low reputations are detected. For malicious Client 1, even though it does not misbehave all the time, it can still be detected and penalized with a low reputation. Fig.~\ref{fig:ma} shows the total number of malicious model updates by malicious clients. After a short while, all these three malicious devices are detected and not able to perform poisoning attacks.

\textbf{Incentive mechanism evaluation.}
We also monitor the accumulative incentives for the same 3 honest clients and 3 malicious clients in the MNIST dataset as in the previous experiment. As shown in Fig.~\ref{fig:rew}, honest clients can always get rewards even if sometimes they are not able to provide good model updates. The honest Client 1 who always provides high-quality updates can gain more rewards. On the other hand, malicious clients might earn some rewards at the beginning, however, after a short period, they will be detected and cannot gain any further profit. The malicious Client 1 behaves correctly sometimes and it is hard to detect. But after round 95, it acquires an extremely low reputation which makes it expelled from the system, and cannot gain rewards.
	\section{Related Work}
\label{sec:related}

\begin{table*}[thbp]
	\centering
        \footnotesize
	\resizebox{0.8\textwidth}{!}{
		\renewcommand\arraystretch{.5}
		\setlength{\tabcolsep}{2mm}
		\begin{tabular}{ccccccc}
			\toprule
			System & Asynchronous & Global model & Dynamic & Incentive & Privacy & Security\\
			\midrule
			BLADE-FL \cite{li2021blockchain}& No & Yes & No & No & Yes & Yes \\
			\midrule
            Biscotti \cite{shayan2020biscotti}& No & Yes & Yes & Yes & Yes & Yes \\
			\midrule
            BAFFLE \cite{ramanan2020baffle}& No & Yes & No & Yes & Not discussed & Not discussed \\
            \midrule
            VFChain \cite{peng2021vfchain}& No & Yes & Yes & Yes & Yes & Yes \\
            \midrule
            BAFL \cite{feng2021bafl}& Yes & Yes & Yes & Yes & No & Yes \\
            \midrule
            \textbf{BDFL (this work)}& Yes & No & Yes & Yes & Yes & Yes\\ 
			\bottomrule
	\end{tabular}}
	\caption{List of Blockchain-based Federated Learning System. }
	\vspace{-3ex}
	\label{table:related}
\end{table*}

\addtolength{\textheight}{-0.09in}

Federated Learning with a centralized aggregator reveals a single point of failure and is vulnerable to malicious clients and false data. Thus, more decentralized setups have been proposed to address these limitations of centralized architecture, such as Blockchain-based Federated Learning.
We review the existing state-of-the-art blockchain-based federated learning systems and summarize them in Table.~\ref{table:related}. 

BLADE-FL~\cite{li2021blockchain} proposed a blockchain-assisted decentralized FL, in which each client broadcasts the trained model to other clients, aggregates its own model with received ones, and then competes to generate a block before its local training of the next round. However, in FL, clients might not have enough computing capability to do both training and mining.
Biscotti~\cite{shayan2020biscotti} focused on the security and privacy issue between peering clients. Instead of doing training and mining at the same time, clients' roles (such as verifier and aggregator) are selected randomly each round. However, it still requires all clients to agree on a global model at the end of each round.
BAFFLE~\cite{ramanan2020baffle} leverages smart contracts to maintain the global model copy and the associated computational state of the users. The machine learning model weight vector is partitioned into numerous chunks to be stored in the smart contracts. However, with the dramatically growing size of the ML models, chunking might become extremely challenging.
VFChain~\cite{peng2021vfchain} utilized blockchain to verify and audit the correctness of the training process for federated learning. Instead of storing all the model updates, it only records the related verifiable information of models in the blockchain for audit in the future. However, same as in previous works, it requires synchronization between all clients.
BAFL~\cite{feng2021bafl} introduced an asynchronous FL framework that uses a blockchain for model aggregation. 
However, every client needs to upload its local model to the blockchain, which is even more inefficient.
In traditional distributed machine learning, there are also works\cite{zhou2021truthful, zhou2022dps, pang2022incentive} concerning incentives for clients.

Compared to existing work, BDFL is the first solution for asynchronous blockchain-based fully decentralized FL, in which clients can join and leave the system dynamically and exchange models in a P2P manner without the need to maintain a global model in each round.

\section{Conclusion}
\label{sec:conclusion}
We present BDFL, a blockchain-based fully decentralized federated learning system that enables clients to exchange models in a P2P manner with model verification with high learning accuracy and system robustness. We design an incentive mechanism to encourage clients to participate in the model exchange, and a reputation model to evaluate the trustworthiness of each client to avoid malicious model updates from attackers.
The evaluation results via simulations show that BDFL achieves fast model convergence and high accuracy on real datasets even if there exist 30\% malicious clients in the system.

\vspace{-2ex}
\addtolength{\textheight}{-0.02in}

 \section{Acknowledgement}
The authors were partially supported by NSF Grants 1750704, 1932447, and 2114113. C. Qian was partially supported by the Army Research Office and was accomplished under Grant Number W911NF-20-1-0253. The views and conclusions contained in this document are those of the authors and should not be interpreted as representing the official policies, either expressed or implied, of the Army Research Office or the U.S. Government. The U.S. Government is authorized to reproduce and distribute reprints for Government purposes notwithstanding any copyright notation herein.
We thank the anonymous reviewers for their comments.

	\bibliographystyle{IEEEtran}
	\bibliography{IEEEabrv,IEEEexample}

\begin{thebibliography}{10}
\providecommand{\url}[1]{#1}
\csname url@samestyle\endcsname
\providecommand{\newblock}{\relax}
\providecommand{\bibinfo}[2]{#2}
\providecommand{\BIBentrySTDinterwordspacing}{\spaceskip=0pt\relax}
\providecommand{\BIBentryALTinterwordstretchfactor}{4}
\providecommand{\BIBentryALTinterwordspacing}{\spaceskip=\fontdimen2\font plus
\BIBentryALTinterwordstretchfactor\fontdimen3\font minus
  \fontdimen4\font\relax}
\providecommand{\BIBforeignlanguage}[2]{{%
\expandafter\ifx\csname l@#1\endcsname\relax
\typeout{** WARNING: IEEEtran.bst: No hyphenation pattern has been}%
\typeout{** loaded for the language `#1'. Using the pattern for}%
\typeout{** the default language instead.}%
\else
\language=\csname l@#1\endcsname
\fi
#2}}
\providecommand{\BIBdecl}{\relax}
\BIBdecl

\bibitem{mcmahan2017communication}
B.~McMahan, E.~Moore, D.~Ramage, S.~Hampson, and B.~A. y~Arcas,
  ``Communication-efficient learning of deep networks from decentralized
  data,'' in \emph{Artificial intelligence and statistics}.\hskip 1em plus
  0.5em minus 0.4em\relax PMLR, 2017.

\bibitem{ma2020safeguarding}
C.~Ma, J.~Li, M.~Ding, H.~H. Yang, F.~Shu, T.~Q. Quek, and H.~V. Poor, ``On
  safeguarding privacy and security in the framework of federated learning,''
  \emph{IEEE network}, 2020.

\bibitem{wei2020federated}
K.~Wei, J.~Li, M.~Ding, C.~Ma, H.~H. Yang, F.~Farokhi, S.~Jin, T.~Q. Quek, and
  H.~V. Poor, ``Federated learning with differential privacy: Algorithms and
  performance analysis,'' \emph{IEEE TIFS}, 2020.

\bibitem{li2020federated}
T.~Li, A.~K. Sahu, A.~Talwalkar, and V.~Smith, ``Federated learning:
  Challenges, methods, and future directions,'' \emph{IEEE signal processing
  magazine}, 2020.

\bibitem{he2018cola}
L.~He, A.~Bian, and M.~Jaggi, ``Cola: Decentralized linear learning,''
  \emph{Advances in NIPS}, 2018.

\bibitem{sun2022decentralized}
T.~Sun, D.~Li, and B.~Wang, ``Decentralized federated averaging,''
  \emph{Proceedings of IEEE TPAMI}, 2022.

\bibitem{Liu2023AAFL}
J.~Liu, H.~Xu, L.~Wang, Y.~Xu, C.~Qian, J.~Huang, and H.~Huang, ``Adaptive
  asynchronous federated learning in resource-constrained edge computing,''
  \emph{IEEE TMC}, 2023.

\bibitem{liao2023ADFL}
Y.~Liao, Y.~Xu, H.~Xu, L.~Wang, and C.~Qian, ``Adaptive configuration for
  heterogeneous participants in decentralized federated learning,'' in
  \emph{Proceedings of IEEE INFOCOM}, 2023.

\bibitem{shayan2020biscotti}
M.~Shayan, C.~Fung, C.~J. Yoon, and I.~Beschastnikh, ``Biscotti: A blockchain
  system for private and secure federated learning,'' \emph{proceedings of IEEE
  TPDS}, 2020.

\bibitem{gervais2016security}
A.~Gervais, G.~O. Karame, K.~W{\"u}st, V.~Glykantzis, H.~Ritzdorf, and
  S.~Capkun, ``On the security and performance of proof of work blockchains,''
  in \emph{Proceedings of ACM CCS}, 2016.

\bibitem{biggio2012poisoning}
B.~Biggio, B.~Nelson, P.~Laskov \emph{et~al.}, ``Poisoning attacks against
  support vector machines,'' in \emph{Proceedings of ICML}, 2012.

\bibitem{fung2020limitations}
C.~Fung, C.~J. Yoon, and I.~Beschastnikh, ``The limitations of federated
  learning in sybil settings.'' in \emph{RAID}, 2020.

\bibitem{hitaj2017deep}
B.~Hitaj, G.~Ateniese, and F.~Perez-Cruz, ``Deep models under the gan:
  information leakage from collaborative deep learning,'' in \emph{Proceedings
  of ACM CCS}, 2017.

\bibitem{salem2018ml}
A.~Salem, Y.~Zhang, M.~Humbert, P.~Berrang, M.~Fritz, and M.~Backes,
  ``Ml-leaks: Model and data independent membership inference attacks and
  defenses on machine learning models,'' \emph{arXiv preprint
  arXiv:1806.01246}, 2018.

\bibitem{abadi2016deep}
M.~Abadi, A.~Chu, I.~Goodfellow, H.~B. McMahan, I.~Mironov, K.~Talwar, and
  L.~Zhang, ``Deep learning with differential privacy,'' in \emph{Proceedings
  of ACM CCS}, 2016.

\bibitem{dwork2014algorithmic}
C.~Dwork, A.~Roth \emph{et~al.}, ``The algorithmic foundations of differential
  privacy,'' \emph{Foundations and Trends{\textregistered} in Theoretical
  Computer Science}, 2014.

\bibitem{zamani2018rapidchain}
M.~Zamani, M.~Movahedi, and M.~Raykova, ``Rapidchain: Scaling blockchain via
  full sharding,'' in \emph{Proceedings of ACM CCS}, 2018.

\bibitem{hua2022efficient}
Y.~Hua, K.~Miller, A.~L. Bertozzi, C.~Qian, and B.~Wang, ``Efficient and
  reliable overlay networks for decentralized federated learning,'' \emph{SIAM
  Journal on Applied Mathematics}, vol.~82, no.~4, pp. 1558--1586, 2022.

\bibitem{yu2014space}
Y.~Yu and C.~Qian, ``Space shuffle: A scalable, flexible, and high-bandwidth
  data center network,'' in \emph{Proceedings of IEEE ICNP}, 2014.

\bibitem{sheller2020federated}
M.~J. Sheller, B.~Edwards, G.~A. Reina, J.~Martin, S.~Pati, A.~Kotrotsou,
  M.~Milchenko, W.~Xu, D.~Marcus, R.~R. Colen \emph{et~al.}, ``Federated
  learning in medicine: facilitating multi-institutional collaborations without
  sharing patient data,'' \emph{Scientific reports}, 2020.

\bibitem{mavroudis2020snappy}
V.~Mavroudis, K.~W{\"u}st, A.~Dhar, K.~Kostiainen, and S.~Capkun, ``Snappy:
  Fast on-chain payments with practical collaterals,'' in \emph{Proceedings of
  USENIX NDSS}, 2020.

\bibitem{castro2002practical}
M.~Castro and B.~Liskov, ``Practical byzantine fault tolerance and proactive
  recovery,'' \emph{ACM TOCS}, 2002.

\bibitem{bell2020secure}
J.~H. Bell, K.~A. Bonawitz, A.~Gasc{\'o}n, T.~Lepoint, and M.~Raykova, ``Secure
  single-server aggregation with (poly) logarithmic overhead,'' in
  \emph{Proceedings of ACM CCS}, 2020.

\bibitem{bonawitz2017practical}
K.~Bonawitz, V.~Ivanov, B.~Kreuter, A.~Marcedone, H.~B. McMahan, S.~Patel,
  D.~Ramage, A.~Segal, and K.~Seth, ``Practical secure aggregation for
  privacy-preserving machine learning,'' in \emph{proceedings of ACM CCS},
  2017.

\bibitem{choudhury2020anonymizing}
O.~Choudhury, A.~Gkoulalas-Divanis, T.~Salonidis, I.~Sylla, Y.~Park, G.~Hsu,
  and A.~Das, ``Anonymizing data for privacy-preserving federated learning,''
  \emph{arXiv preprint arXiv:2002.09096}, 2020.

\bibitem{fang2020local}
M.~Fang, X.~Cao, J.~Jia, and N.~Z. Gong, ``Local model poisoning attacks to
  byzantine-robust federated learning,'' in \emph{Proceedings of USENIX
  Security}, 2020.

\bibitem{dwork2006our}
C.~Dwork, K.~Kenthapadi, F.~McSherry, I.~Mironov, and M.~Naor, ``Our data,
  ourselves: Privacy via distributed noise generation,'' in
  \emph{EUROCRYPT}.\hskip 1em plus 0.5em minus 0.4em\relax Springer, 2006.

\bibitem{peng2021vfchain}
Z.~Peng, J.~Xu, X.~Chu, S.~Gao, Y.~Yao, R.~Gu, and Y.~Tang, ``Vfchain: Enabling
  verifiable and auditable federated learning via blockchain systems,''
  \emph{IEEE TNSE}, 2021.

\bibitem{deng2012mnist}
L.~Deng, ``The mnist database of handwritten digit images for machine learning
  research [best of the web],'' \emph{IEEE signal processing magazine}, 2012.

\bibitem{krizhevsky2009learning}
A.~Krizhevsky, G.~Hinton \emph{et~al.}, ``Learning multiple layers of features
  from tiny images,'' 2009.

\bibitem{li2021blockchain}
J.~Li, Y.~Shao, K.~Wei, M.~Ding, C.~Ma, L.~Shi, Z.~Han, and H.~V. Poor,
  ``Blockchain assisted decentralized federated learning (blade-fl):
  Performance analysis and resource allocation,'' \emph{IEEE Transactions on
  Parallel and Distributed Systems}, 2021.

\bibitem{ramanan2020baffle}
P.~Ramanan and K.~Nakayama, ``Baffle: Blockchain based aggregator free
  federated learning,'' in \emph{IEEE international conference on blockchain
  (Blockchain)}, 2020.

\bibitem{feng2021bafl}
L.~Feng, Y.~Zhao, S.~Guo, X.~Qiu, W.~Li, and P.~Yu, ``Bafl: A blockchain-based
  asynchronous federated learning framework,'' \emph{IEEE Transactions on
  Computers}, 2021.

\bibitem{zhou2021truthful}
R.~Zhou, J.~Pang, Z.~Wang, J.~C. Lui, and Z.~Li, ``A truthful procurement
  auction for incentivizing heterogeneous clients in federated learning,'' in
  \emph{2021 IEEE 41st International Conference on Distributed Computing
  Systems (ICDCS)}.\hskip 1em plus 0.5em minus 0.4em\relax IEEE, 2021, pp.
  183--193.

\bibitem{zhou2022dps}
R.~Zhou, N.~Wang, Y.~Huang, J.~Pang, and H.~Chen, ``Dps: Dynamic pricing and
  scheduling for distributed machine learning jobs in edge-cloud networks,''
  \emph{IEEE Transactions on Mobile Computing}, 2022.

\bibitem{pang2022incentive}
J.~Pang, J.~Yu, R.~Zhou, and J.~C. Lui, ``An incentive auction for
  heterogeneous client selection in federated learning,'' \emph{IEEE
  Transactions on Mobile Computing}, 2022.

\end{thebibliography}

	%
\end{document}